\documentclass[a4paper]{article}
\usepackage{listings}
\usepackage{xcolor}
\usepackage{multirow}
\usepackage{jheppub}
\usepackage{dcolumn}
\usepackage[english]{babel}
\usepackage[utf8x]{inputenc}
\usepackage[T1]{fontenc}
\usepackage{palatino}
\usepackage{amsmath}
\usepackage{afterpage}
\usepackage{graphicx, subcaption}
\usepackage[colorinlistoftodos]{todonotes}
\usepackage[colorlinks=true]{hyperref}
\usepackage{makeidx}
\newcommand{\be}{\begin{equation}}
\newcommand{\ee}{\end{equation}}
\newcommand{\bea}{\begin{eqnarray}}
\newcommand{\eea}{\end{eqnarray}}
\makeindex
\begin{document}
\vspace*{1.2cm}

\thispagestyle{empty}
\begin{center}
{\LARGE \bf Charmonia photo-production in ultra-peripheral and peripheral PbPb collisions with LHCb}

\par\vspace*{7mm}\par

{

\bigskip

\large \bf Weisong Duan on behalf of the LHCb collaboration}

\bigskip

{\large \bf  E-Mail: weisong.duan@cern.ch}

\bigskip

{South China Normal University, Guangzhou, China}

\bigskip

{\it Presented at the Low-$x$ Workshop, Elba Island, Italy, September 27--October 1 2021}

\vspace*{15mm}

\end{center}
\vspace*{1mm}

\begin{abstract}

The LHCb recorded ∼ 210 $\mu b^{-1}$ integrated luminosity of PbPb collisions at $\sqrt{s_{NN}}$ = 5.02 TeV in 2018. With an increase of the luminosity by a factor of 20 compared to the previous 2015 PbPb dataset, precise measurements on photo-produced charmonia in ultra-peripheral collisions are now possible. Moreover, the great momentum resolution of the detector allows photo-produced $J/\psi$ in collisions with a nuclear overlap to be studied. This new type of probe is sensitive to the geometry of the collisions but also to the electromagnetic field of the Pb nuclei. In this contribution, we present the latest results on $J/\psi$ photo-production measured by LHCb in peripheral and ultra-peripheral PbPb collisions.
\end{abstract}
 
\section{Introduction}
The LHCb detector is a single-arm forward spectrometer fully instrumented in the pseudorapidity range 2 < $\eta$ < 5~\cite{Collaboration_2008}. It has a high precision tracking system, which provides excellent vertex and momentum resolution, and full particle identification. Compared to the ALICE, CMS and ATLAS, LHCb covers the forward rapidity region, providing better access to the gluon distribution at small $x$.

The photonuclear production of vector mesons such as $J/\psi$ is sensitive to the gluon parton distribution function in the nucleus at small Bjorken-$x$, which is estimated by $x \approx (m_{J/\psi}\cdot e^{-y})/\sqrt{s_{NN}}$, where $m_{J/\psi}$ is mass of $J/\psi$ and $y$ is its rapidity. Coherent photoproduction of $J/\psi$ meson provides a way to study the nuclear shadowing effects at small Bjorken-$x$ ranging from $10^{-5}$ to $10^{-2}$ at LHC energies.
\section{Study of coherent $J/\psi$ production in ultra-peripheral lead-lead collisions at $\sqrt{s_{NN}}$ = 5 TeV}

The ultra-peripheral collisions, UPCs, are $\rm{PbPb \to Pb + Pb + X}$ in which two ions interact via their cloud of virtual photons. If the photon couples coherently to the nucleus as a whole, it is called coherent production. If the photon couples with one nucleon leading to the breakup of the target nucleus, it is called incoherent production.

In UPCs, coherent $J/\psi$ meson production can be described by the interaction between photons and gluons, according to the Regge theory~\cite{PhysRevC.86.014905, PhysRevC.93.055206}, gluons are considered as a single object with vacuum quantum numbers, which is called pomeron ($\enspace \rm P \kern -1em I$ $\enspace$). The cross-section for photoproduction gives constraints on the gluon parton distribution functions. This process has low multiplicities and very low transverse momentum $p_{T}$.

The $J/\psi$ mesons are reconstructed through the $J/\psi \to \mu^{+}\mu^{-}$ decay channel, using 2015 Pb-Pb data samples, corresponding to an integrated luminosity of 10 $\mu b^{-1}$. An ultra-peripheral electromagnetic interaction could occur simultaneously with the hadron collision. The HeRSCheL detector~\cite{Akiba_2018} is used to reject backgrounds from hadronic interactions.

\begin{figure}[htbp]
\begin{center}
\epsfig{figure=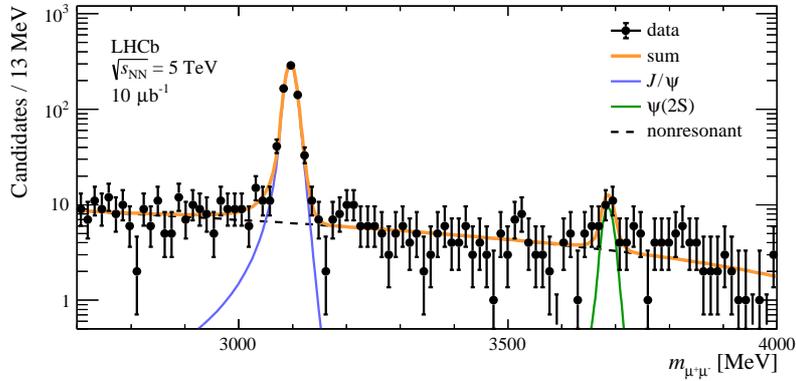,height=0.35\textwidth}
\caption{The di-muon invariant mass distribution in the rapidity range between 2.0 and 4.5. The solid blue line corresponds to the $J/\psi$ meson. The solid green line represents the $\psi(2S)$ meson. The dashed black line represents non-resonant background.}
\label{massfit}
\end{center}
\end{figure}

The number of candidates are obtained by fitting the di-muon spectrum as shown in Fig.~\ref{massfit}. The $J/\psi$ and $\psi(2S)$ mass spectrum are modeled by a double-sided Crystal ball function, and the non-resonant background is modeled by exponential function multiplied by an first-order polynomial function. Thus we determine the number of $J/\psi$ candidates within the $J/\psi$ mass window 3040 MeV/$c^{2}$ $\sim$ 3165 MeV/$c^{2}$ and the number of $\psi(2S)$ candidates within the $\psi(2S)$ mass window 3608 MeV/$c^{2}$ $\sim$ 3763 MeV/$c^{2}$.

\begin{figure}[htbp]
\begin{center}
\epsfig{figure=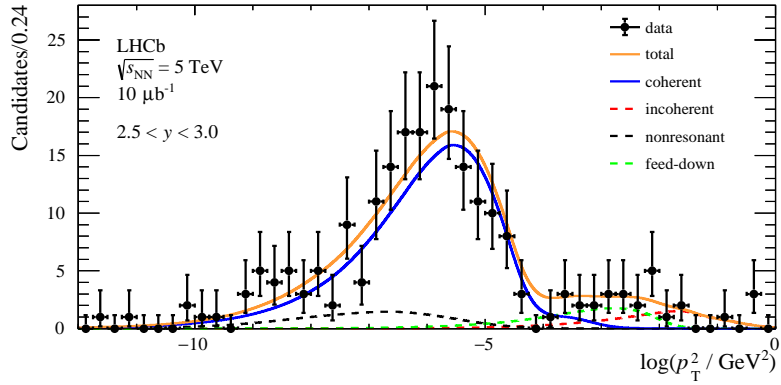,height=0.35\textwidth}
\caption{The ${\rm log} (p^{2}_{T})$ distribution of $J/\psi$ candidates in the interval 2.5 < $y$ < 3.0, where the unit of $p_{T}$ is GeV/c. The solid blue line represents coherent $J/\psi$ distribution. The dashed red line represents the incoherent distribution. The dashed green line represents feed down from $\psi(2S)$. The dashed black line represents non-resonant background.}
\label{logpt2fit}
\end{center}
\end{figure}

To determine the coherent $J/\psi$ production, a fit to the ${\rm log} (p^{2}_{T})$ is performed to extract the coherent $J/\psi$ mesons within the $J/\psi$ mass window. The ${\rm log} (p^{2}_{T})$ distribution of the $J/\psi$ mesons is shown in Fig.~\ref{logpt2fit}. In short, the number of inclusive $J/\psi$ mesons is obtained by the invariant mass fit, and the number of coherent $J/\psi$ mesons is obtained by the ${\rm log} (p^{2}_{T})$ fit.

\begin{figure}
\begin{center}
\epsfig{figure=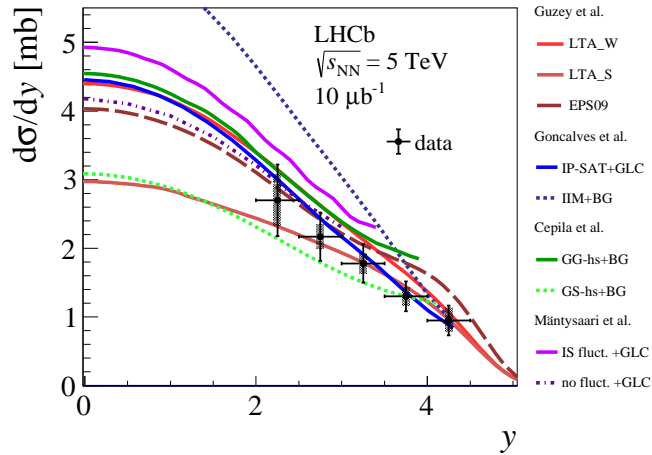,height=0.45\textwidth}
\caption{Differential cross-section of coherent $J/\psi$ production as a function of rapidity, with comparisons to phenomenological models.}
\label{csresults}
\end{center}
\end{figure}

The results of the differential cross section are calculated in five rapidity bins as shown in Fig.~\ref{csresults}.  We compared the results between experimental results and theoretical predictions~\cite{PhysRevC.93.055206, PhysRevC.84.011902, PhysRevC.97.024901, PhysRevD.96.094027, MANTYSAARI2017832}. The coherent $J/\psi$ cross-section production is given by:

\begin{equation}
    \frac{\mathrm{d}\sigma_{coh,J/\psi}}{\mathrm{d}y} = \frac{N_{coh,J/\psi}}{\varepsilon_{t}\cdot \mathcal{L}\cdot\Delta y \cdot \mathcal{B}(J/\psi \to \mu^{+}\mu^{-})}
\end{equation}

where $\varepsilon_{t}$ is the total efficiency, $\mathcal{L}$ is an integrated luminosity of the Pb-Pb data sample, and $\mathcal{B} = (5.961 ± 0.033)\%$ is the $J/\psi \to \mu^{+}\mu^{-}$ branching ratio.

\begin{figure}
\begin{center}
\epsfig{figure=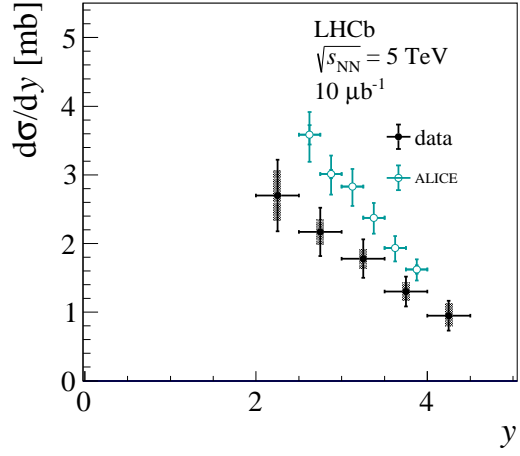,height=0.45\textwidth}
\caption{Differential cross-section of coherent $J/\psi$ production as a function of rapidity, with comparison to ALICE measurements.}
\label{cscompare}
\end{center}
\end{figure}

\begin{figure}
\begin{center}
\epsfig{figure=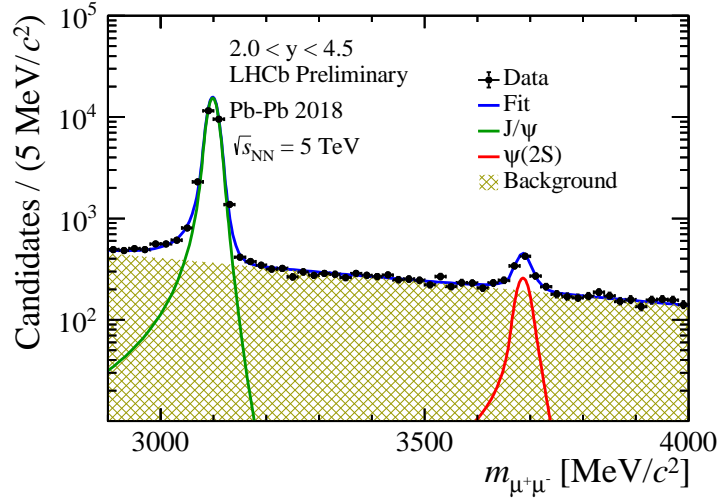,height=0.45\textwidth}
\caption{The invariant mass distribution of $J/\psi$ and $\psi(2S)$ candidates. The solid green line represents $J/\psi$ signal, the solid red line represents $\psi(2S)$ signal, and the yellow regions represents non-resonant background.}
\label{2018massfit}
\end{center}
\end{figure}

Similarly, ALICE measured the coherent $J/\psi$ production cross-section~\cite{2019134926}, so we also compared the coherent $J/\psi$ production cross-section between ALICE and LHCb, as shown in Fig.~\ref{cscompare}. The LHCb result is slightly lower than the ALICE measurement by around 1.3 $\sigma$. Measurements of the coherent $J/\psi$ and $\psi(2S)$ are currently underway using 2018 Pb-Pb data, corresponding to an integrated luminosity of 210 $\mu b^{-1}$. Fig.~\ref{2018massfit} shows the di-muon invariant mass distribution in the range between 2.7 and 4.0 GeV. The final results are expected in the near future.

\section{Study of $J/\psi$ photo-production in lead-lead peripheral collisions at $\sqrt{s_{NN}}$ = 5 TeV}

The second result in this contribution is photo-production of $J/\psi$ at low $p_{T}$, studied in  peripheral Pb-Pb collisions at $\sqrt{s_{NN}}$ = 5 TeV, using the data sample collected by LHCb in 2018, with an integrated luminosity of about 210 $\mu b^{-1}$~\cite{lhcbcollaboration2021study}. 

\begin{figure}[htbp]
\begin{center}
\epsfig{figure=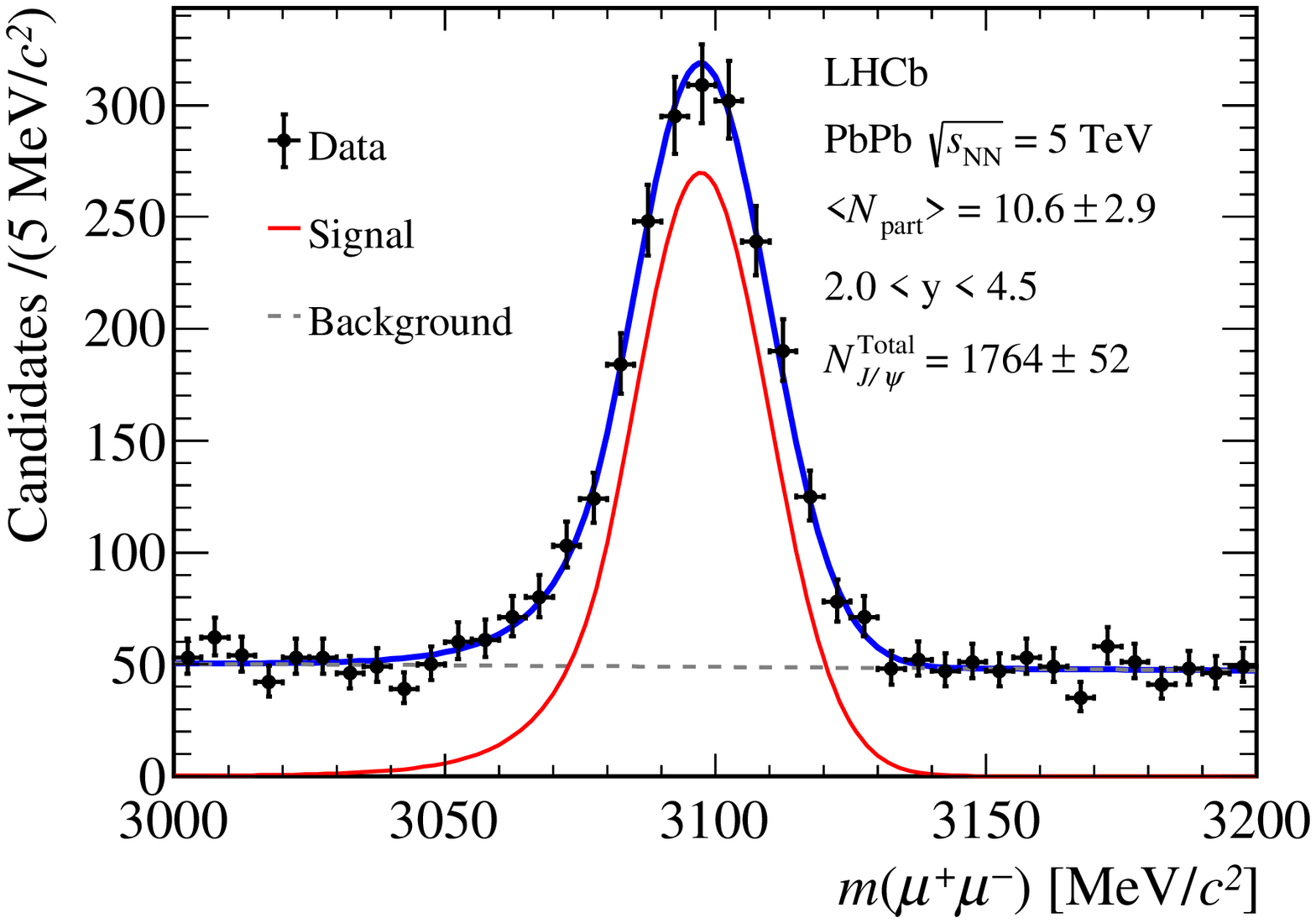,height=0.33\textwidth}
\epsfig{figure=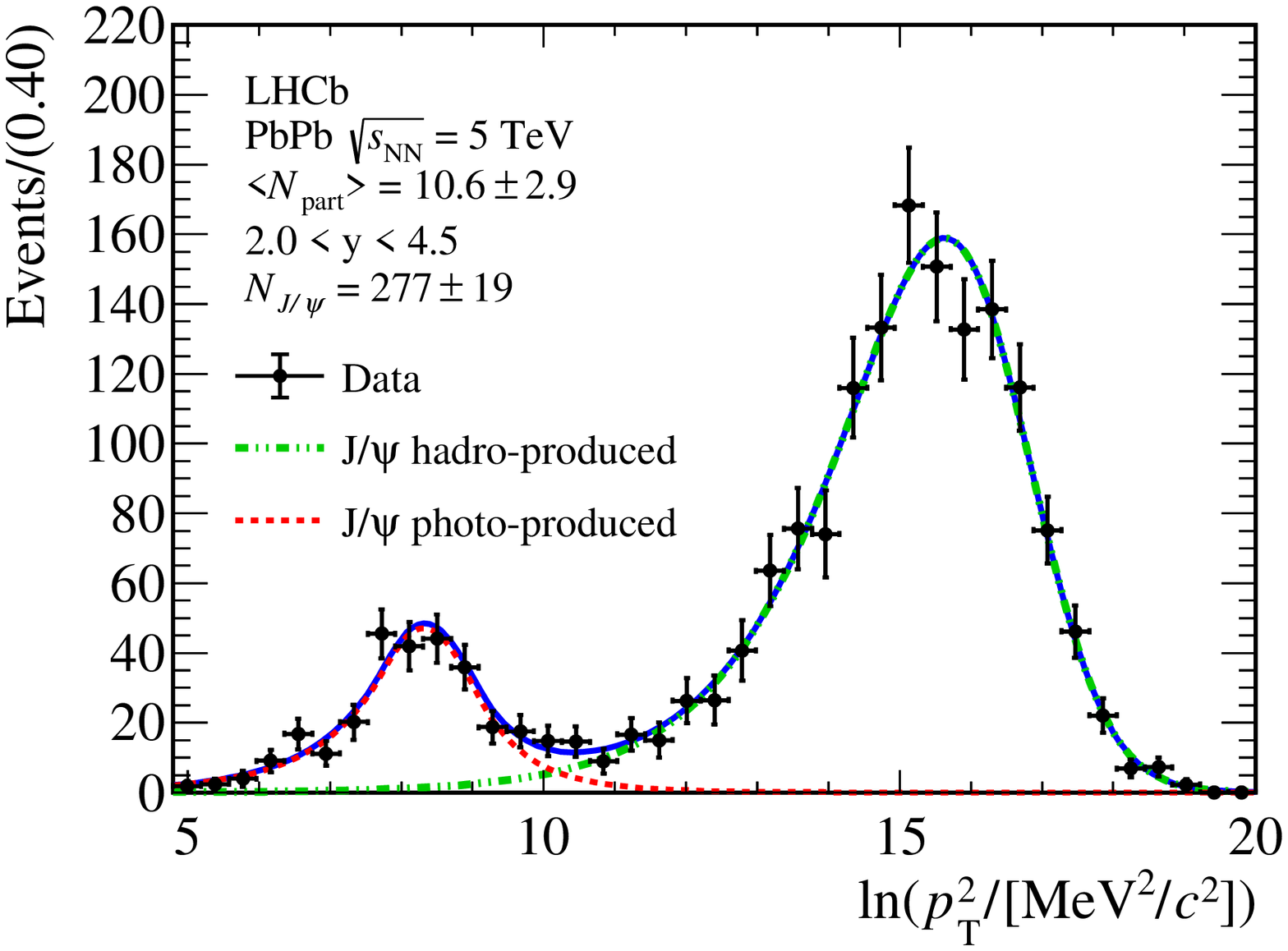,height=0.33\textwidth}
\caption{The left plot is the invariant mass distribution of $J/\psi$ candidates in $p_{T}$ < 15.0 GeV/c and 2 < $y$ <4.5, with $\left<N_{part}\right>$ = 10.6 $\pm$ 2.9. The right plot is the ${\rm ln} (p^{2}_{T})$ distribution of $J/\psi$ candidates after background subtraction for the same kinematic interval.}
\label{mass_and_pt}
\end{center}
\end{figure}

The $J/\psi$ candidates are selected through the $J/\psi \to \mu^{+}\mu^{-}$ decay channel. The di-muon invariant mass spectrum of the selected candidates in the range between 3.0 and 3.2 GeV is shown in the left of Fig.~\ref{mass_and_pt}, for the $J/\psi$ mesons with $\rm{p_{T}}$ <15.0 GeV/c and the number of participants $\left<N_{part}\right>$ = 10.6 $\pm$ 2.9 in full LHCb rapidity coverage 2 < $y$ <4.5. 
 
The inclusive $J/\psi$ candidates consists of photo-produced and hadronically produced $J/\psi$ mesons, which are separated by an unbinned  maximum likelihood fit to the ${\rm log} (p^{2}_{T})$ distribution, as shown in the right of Fig.~\ref{mass_and_pt}. In this figure, the transverse momentum of photo-produced $J/\psi$ yields (red dotted line) are visible in the range between 0 and 250 MeV/c.

\begin{figure}[htbp]
\begin{center}
\epsfig{figure=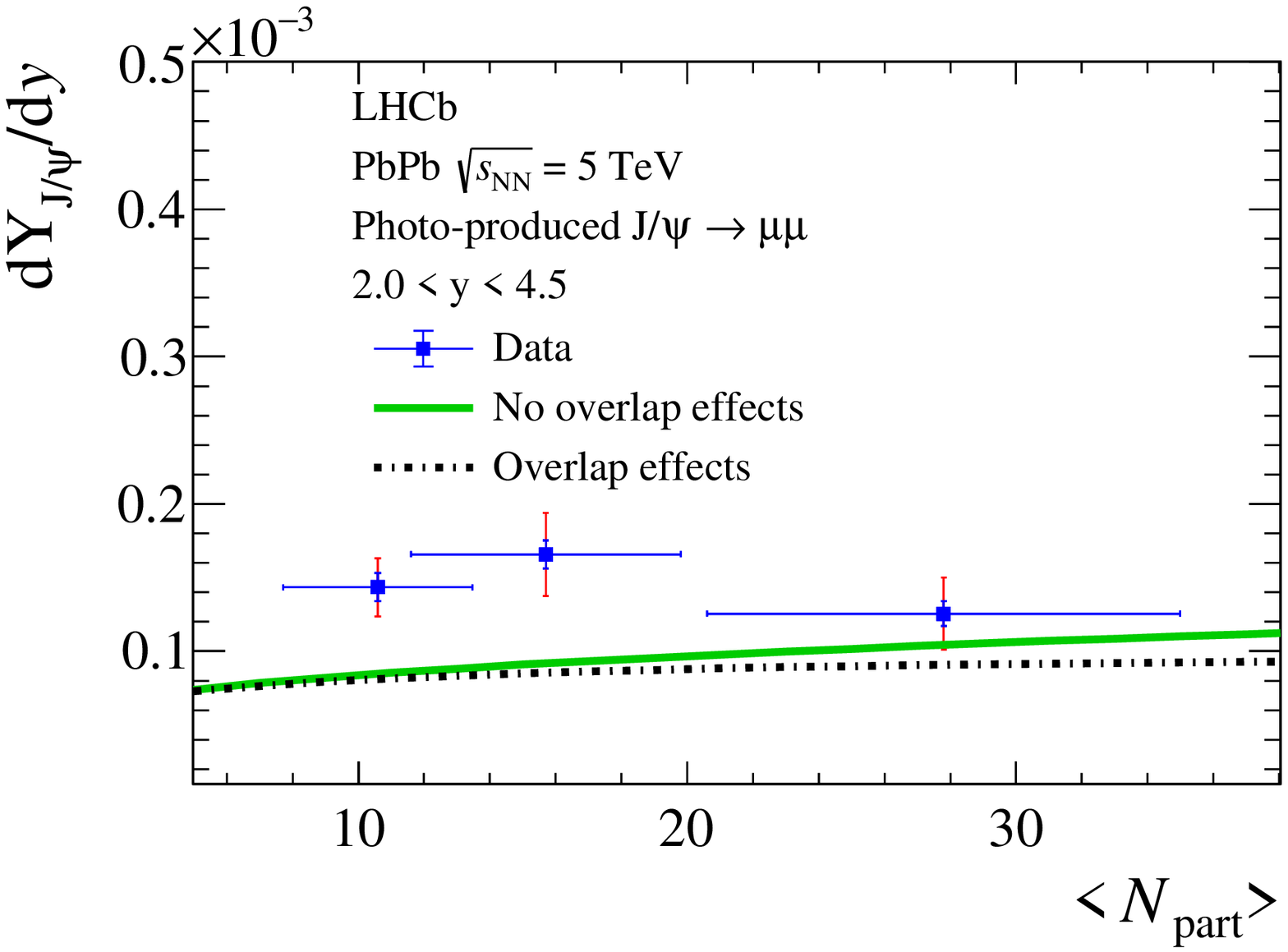,height=0.33\textwidth}
\epsfig{figure=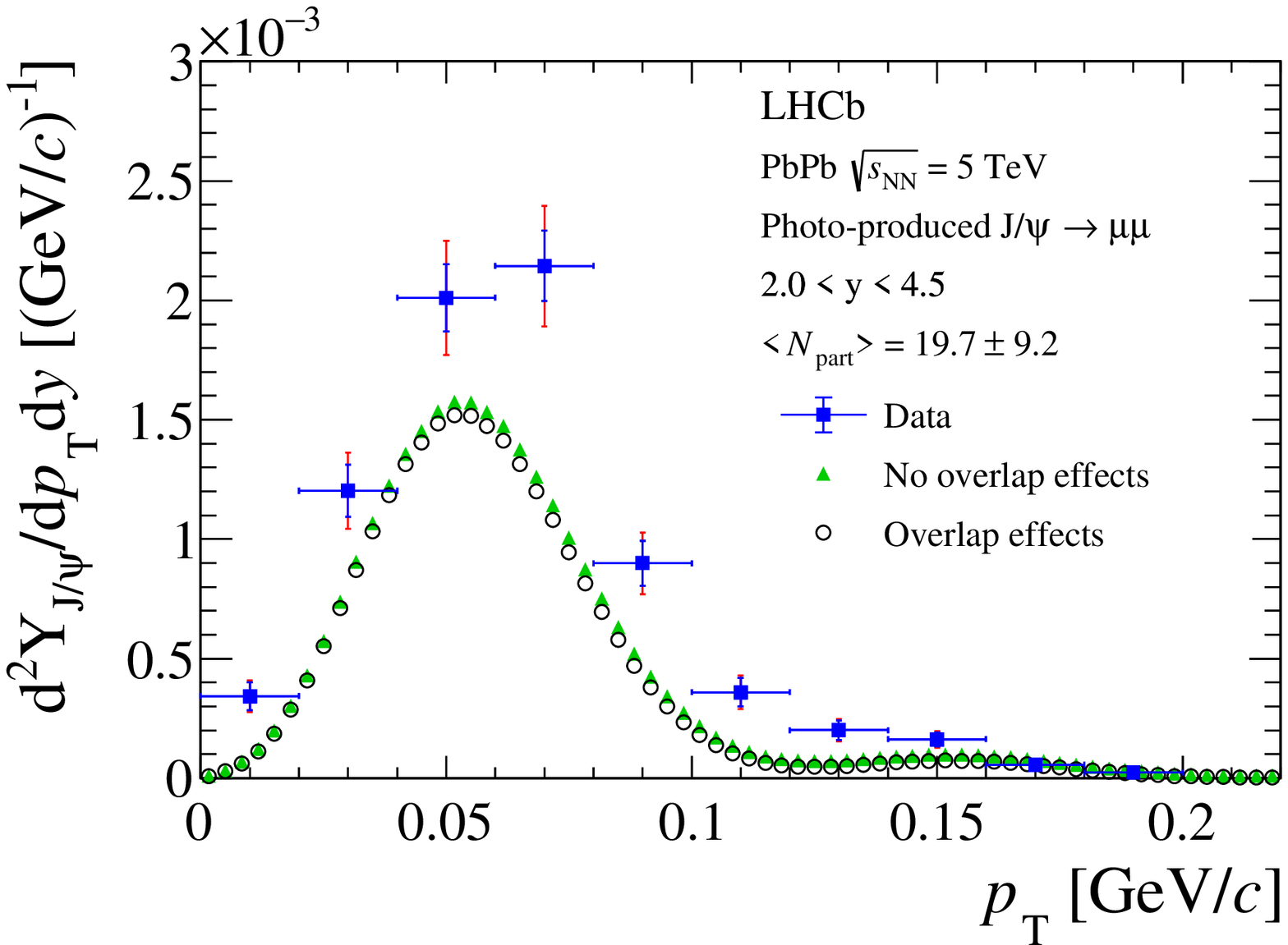,height=0.33\textwidth}
\caption{Illustration of invariant yields of photo-produced $J/\psi$ mesons as a function of $N_{part}$ and $p_{T}$. Note that the blue error bars represent the statistical uncertainty and the red error bars the total uncertainty. Theoretical model predictions~\cite{PhysRevC.97.044910, PhysRevC.99.061901} are shown in open circles in black and green.}
\label{compare}
\end{center}
\end{figure}

Fig.~\ref{compare} shows the photo-produced $J/\psi$ meson yields as a function of $p_{T}$(right), and $\left<N_{part}\right>$ (left). The mean $p_{T}$ of the coherent $J/\psi$ is estimated to be $\left<p_{T}\right>$ = 64.9 $\pm$ 2.4 MeV/c. Theoretical predictions~\cite{PhysRevC.97.044910, PhysRevC.99.061901} are drawn in open circles, and are qualitatively in agreement with the experimental results in shape.

\section{Conclusions}

LHCb detector's unique geometry acceptance allows us to study the nuclear shadowing in small $x$ region through the UPC collisions. The $J/\psi$ photoproduction in UPC using PbPb collisions at 5TeV is measured by LHCb, and is compared to results by ALICE. The photoproduced $J/\psi$ production in peripheral PbPb collisions is also measured with high precision at very low $p_{T}$, and is compared to theoretical calculations. These results demonstrate the capabilities of the LHCb detector in studying nuclear effects. More results from the large 2018 PbPb dataset are expected in the future.

\end{document}